\documentclass[12pt]{iopart}
\usepackage{graphicx}
\usepackage{cite}
\begin{document}
\title{Simulations for trapping reactions with subdiffusive traps
and subdiffusive particles}
\author{J. J. Ruiz-Lorenzo$^{1,2}$, Santos B. Yuste$^{1}$ and Katja
Lindenberg$^{3}$}
\address{$^{(1)}$ Departamento de F\'{\i}sica, Universidad de
Extremadura, E-06071 Badajoz, Spain\\
$^{(2)}$ Instituto de Biocomputaci\'on y F\'{\i}sica de los Sistemas
Complejos (BIFI), Zaragoza 50009, Spain\\
$^{(3)}$ Department of Chemistry and Biochemistry 0340, and Institute
for
Nonlinear Science,
University of California San Diego, 9500 Gilman Drive, La Jolla, CA
92093-0340, USA}

\begin{abstract}
While there are many well-known and extensively tested results involving
diffusion-limited binary reactions, reactions involving subdiffusive
reactant species are far less understood.  Subdiffusive motion is
characterized by a mean square displacement
$\langle x^2\rangle \sim t^\gamma$ with $0<\gamma<1$.
Recently we calculated the asymptotic survival
probability $P(t)$ of a (sub)diffusive particle ($\gamma^\prime$)
surrounded by (sub)diffusive traps ($\gamma$) in one dimension.   These
are among the few known results for reactions involving species
characterized by different anomalous exponents.  Our results were
obtained by bounding, above and below, the exact survival probability by
two other probabilities that are asymptotically identical (except when
$\gamma^\prime=1$ and $0<\gamma<2/3$). Using this approach, we were not
able to estimate the time of validity of the asymptotic result, nor
the way in which the survival probability approaches this regime.  Toward
this goal, here we present a detailed comparison of the asymptotic
results with numerical simulations.  In some parameter ranges the
asymptotic theory describes the simulation results very well even
for relatively short times.  However, in other regimes more time is
required for the simulation results to approach asymptotic behavior, and
we arrive at situations where we are not able to reach asymptotia
within our computational means.  This is regrettably the case for
$\gamma^\prime=1$ and $0<\gamma<2/3$, where we are therefore not able to
prove or disprove even conjectures about the asymptotic survival probability
of the particle.

\end{abstract}


\submitto{\JPCM}
\maketitle


\section{Introduction}
The survival probability of a particle diffusing in a
one-dimensional medium of diffusive traps has only recently been
calculated (but only
asymptotically)~\cite{bray1,blythe,oshanin,moreau,bray2}.
This is surprising in view of its
long history~\cite{bramson1,bramson2,burlatsky} and that
of its antecedents, the so-called trapping
problem~\cite{hughes1,hughes2,weiss1,weiss2,havlin,benavraham,zumofen,klafter,blumen,redner},
in which the traps are static, and the target
problem~\cite{zumofen,klafter,blumen,redner},
in which the particle does not move.  The antecedent systems could
be translated to tractable
boundary value problems, which is not possible when both
particle and traps move.  The solution is an elegant
``tour de force" in which the desired survival probability is bounded
above and below by two others that can be posed as boundary value
problems and that converge to one another asymptotically.

Recently, we undertook the generalization of the bounding approach to
the case of a subdiffusive particle surrounded by a distribution of
subdiffusive traps~\cite{our1,our2}. Subdiffusion of a
particle is usually characterized
by the time dependence of the mean square of the particle displacement
$x(t)$,
\begin{equation}
\left< x^2(t)\right> \sim \frac{2K_\gamma}{\Gamma(1+\gamma)}
t^\gamma . \label{meansquaredispl}
\end{equation}
Here $K_\gamma$ is the (generalized) diffusion constant, and
$\gamma$ is the exponent that characterizes normal ($\gamma=1$) or
anomalous ($\gamma\neq 1$) diffusion.  In particular,
the process is diffusive when $\gamma=1$ and sudiffusive when $0<\gamma<1$.
There are a variety of models and physical circumstances that
lead to subdiffusion in the trapping and, more generally, in the binary
reaction
context~\cite{zumofen,klafter,blumen,sung1,seki,henry1,henry2,vlad,fedotov,sung2,yuste1,yuste2,yuste3,yuste4}.
Many are based on the continuous time random walk
formalism, where particles are thought of as random walkers with
waiting-time distributions between steps that have broad long-time tails
and consequently infinite moments, $\psi(t)\sim
t^{-1-\gamma}$.
Our work is based on the fractional diffusion equation, which describes
the evolution of the probability density $P(x,t)$
of finding the particle at position $x$ at time $t$ by means of
the fractional partial differential equation (in one
dimension)~\cite{metzler,schneider},
\begin{equation}
\frac{\partial }{\partial t} P(x,t)= K_\gamma
~_{0}D_{t}^{1-\gamma } \frac{\partial^2}{\partial x^2} P(x,t),
\label{Pfracdifu}
\end{equation}
where $K_\gamma$ is the generalized diffusion coefficient that
appears in equation~(\ref{meansquaredispl}), and
$~_{0}\,D_{t}^{1-\gamma } $ is the Riemann-Liouville
operator,
\begin{equation}
~_{0}D_{t}^{1-\gamma } P(x,t)=\frac{1}{\Gamma(\gamma)}
\frac{\partial}{\partial t} \int_0^t d\tau
\frac{P(x,\tau)}{(t-\tau)^{1-\gamma}}.
\end{equation}
The connection between these two approaches is in itself an interesting
subject, see e.g.~\cite{barkai,mainardi}.

The survival probability of a particle $A$ characterized by exponent
$\gamma^\prime$ and generalized diffusion coefficient
$K_{\gamma^\prime}$ surrounded by traps characterized by $\gamma$ and
$K_\gamma$ is bounded as follows~\cite{our1,our2}.  An upper bound
is obtained by
forcing particle $A$ to remain still.  The ``Pascal principle" that says
that the best survival strategy for the particle is to stand still was
proved for the diffusive case in~\cite{burlatsky,moreau,bray2} and for
the subdiffusive problem in our work. The solution of the fractional
subdiffusion equation for the $B$ traps with the location of $A$ as
an appropriate boundary then leads to the upper bound for the survival
probability of $A$,
\begin{equation}
P_U(t)= \exp\left[- 2\rho\frac{\sqrt{K_\gamma t^\gamma}}{\Gamma
\left(1+\frac{\gamma}{2}\right)}\right],
\label{upperresult}
\end{equation}
where $\rho$ is the density of traps.
A lower bound is calculated by allowing particle $A$ to move within a
box of size ${\mathcal L}$ while the traps $B$ are forced to remain
outside of this box.  The box size is then found so as to maximize this
lower bound, with the $t\to\infty$ result
\begin{eqnarray}
P_{L}(t)&=& \frac{e^{-2}}{8\Gamma(1-\gamma')}
\left(\frac{2}{\rho}\right)^2 \frac{1}{K'_{\gamma'} t^{\gamma'}
}\nonumber\\
&&\times \exp
\left[- 2\rho\frac{\sqrt{K_\gamma t^\gamma}}{\Gamma
\left(1+\frac{\gamma}{2}\right)}\right]
\times \left[1+O\left(\frac{1}{\rho^2K'_{\gamma'}
t^{\gamma'} }\right)\right]
\label{eq:bound1}
\end{eqnarray}
for $0<\gamma'<1$.  For the diffusive case ($\gamma'=1$)
\begin{equation}\label{eq:dominant}
P_{L}(t) = \frac{4}{\pi} \exp\left[-2\rho\frac{2\sqrt{K_\gamma
t^\gamma}}{\Gamma(1+\gamma/2)} -3(\pi^2 \rho^2 D't/4)^{1/3}\right].
\end{equation}

With these bounds we see that for a subdiffusive particle ($0<\gamma'<
1$) and diffusive or subdiffusive traps ($0<\gamma \leq 1$) the upper
and lower bounds converge asymptotically (viz., compare the logarithms
of both), so that we arrive at the
explicit asymptotic survival probability
\begin{equation}\label{Ptgral}
P(t)\sim
\exp\left[- 2\rho\frac{\sqrt{K_\gamma t^\gamma}}{\Gamma
\left(1+\frac{\gamma}{2}\right)}\right].
\end{equation}
This result elicits a comment about the so-called ``subordination
principle"~\cite{blumen}, according to which in some cases
asymptotic anomalous
diffusion behavior can be found from corresponding results for
normal diffusion with the
simple replacement of $t$ by $t^\gamma$. This can be understood from
a continuous time random walk perpective because
the average number of jumps made by a subdiffusive walker up to
time $t$
scales as $\langle n \rangle \sim t^\gamma$, and in many instances the
number of jumps is the relevant factor that explains the
behavior of the system.  However, for
systems where each species has a \emph{different} anomalous diffusion
exponent, such a replacement becomes ambiguous. The
result~(\ref{Ptgral}) indicates a subordination principle at work
\emph{as determined by the traps}.  In other words, it is the motion of
the traps that regulates the survival probability of the particle
whether or not the particle moves, provided it does not move ``too
easily,"
i.e., provided it is subdiffusive ($\gamma'<1$).

When the particle is diffusive ($\gamma'=1$) the situation is more
complicated, because its asymptotic survival probability is no longer
necessarily the same as it would be if it stood still.
If $2/3<\gamma<1$, i.e., if the traps move sufficiently easily, the
upper and lower bounds still
converge and Eq.~(\ref{Ptgral}) still holds, that
is, it is still the motion of the traps that determines the asymptotic
survival probability of the particle.  If the traps are
subdiffusive with $\gamma=2/3$ (marginal case), the bounds lead only to a
prediction of the asymptotic time dependence but not of the accompanying
exponential prefactor, i.e., the bounds establish
that $P(t) \sim \exp(-\lambda
t^{1/3})$ but are not able to determine $\lambda$.
In particular, we can not determine whether whether $\lambda$ is given
by the coefficient $2\rho \sqrt{K_{2/3}}/\Gamma(4/3)$
of $t^{1/3}$ in the exponent of Eq.~(\ref{Ptgral}) when
$\gamma=2/3$, which one might conjecture.  Finally, when the
particle is diffusive and the traps are sufficiently slow
($0<\gamma<2/3$), the upper and lower bounds do not have the same
asymptotic time dependence, so we are not able to even functionally bound
the survival probability.  While it is still possible in principle that
the trap-driven subordination principle continues to apply in this regime
so that $\theta=\gamma/2$, we have not been able to prove or disprove
such a conjecture (and the behavior at $\gamma=0$ would
not fall within this conjecture, see below).
If this subordination result is invalid,
it would imply (and would not be surprising) that it is no
longer possible to assume the particle to be standing still, and/or that
it may no longer be (or only be) the exponent of the traps that
regulates the motion.  It is interesting to note that for
$\gamma'=1$ and $\gamma=0$, the traditional ``trapping problem," the
asymptotic survival probability is~\cite{donsker} $P(t)\sim
\exp[-3(\pi^2\rho^2D/4)^{1/3}t^{1/3}]$. One might thus be tempted to
conjecture a behavior of the form $P(t) \sim \exp(-\lambda(\gamma,\rho)
t^{1/3})$ throughout the range $0<\gamma\leq 2/3$, i.e., the exponential
prefactor $\lambda$ would have to depend on $\gamma$ and on $\rho$.
However, we have not been able to prove or disprove this conjecture
either.

This analysis therefore leaves open two important questions, which we
attempt to answer by way of detailed numerical simulations (although we
do not entirely succeed):
\begin{enumerate}
\item
In the cases where the bounds converge asymptotically, how much time
does it take for the result to adequately describe the survival
probability of the particle?  In other words, how rapidly do the upper
and lower bounds converge to one another?
\item
Is it possible to find the asymptotic survival probability for the
cases in which our analysis fails to provide converging bounds?
\end{enumerate}
In part the success or failure of this attempt is of course constrained
by the numerical resources at our disposal; the
difficulties in numerically reaching asymptotia even in diffusion
problems are well known~\cite{grassberger}.

The rest of this paper thus consists mostly of figures and a table presenting
numerical simulation results.  Our purpose is to ascertain the behavior
of $\lambda$ and $\theta$ in the expression
\begin{equation}
P(t) = \exp (-\lambda t^\theta)
\label{expression}
\end{equation}
if indeed we arrive at a regime where the survival probability exhibits
this behavior.
If asymptotia has been reached, we expect both to be constant with time.
If they are constant, we compare their values with those obtained from
the asymptotic result~(\ref{Ptgral}).
To test $\theta$, we plot $\ln[-\ln P(t)]$ vs $\ln
t$.  For some cases where $\theta$ is clearly determined as a result of our
simulations (i.e., constant in time and independent of $\rho$),
we test $\lambda$ by plotting $-\ln P(t)/\chi$ vs $\chi$,
\begin{equation}
\chi \equiv \rho \langle x^2\rangle ^{1/2} = \rho \left(
\frac{2K_\gamma}{\Gamma(1+\gamma)}\right)^{1/2} t^{\gamma/2}.
\end{equation}
$\chi$ is in effect a convenient dimensionless measure of time, and
is also the ratio of the root mean square displacement of a
particle at time $t$ to the average distance $\rho^{-1}$ between traps.

In section~\ref{simulations} we briefly outline our simulation
methodology.  Section~\ref{results} is a compendium of
our results, along with the associated descriptions.  A
recapitulation is presented in section~\ref{conclusions}.

\section{Numerical simulation methodology}
\label{simulations}
A brief review of our numerical simulation methodology is appropriate at
this point. We generate the trap distribution by placing a trap at each
site of a
one-dimensional lattice with probability $\rho$ (and not placing a trap
with probability $1-\rho$).  The particle $A$ is placed at
the origin of the lattice. The typical lattice has $10\,000$ sites, and
we implement periodic boundary conditions. We have simulated
larger lattices and different (free) boundary conditions to ascertain
that the results are not affected.

The dynamics of a moving particle in a sea of moving traps is
implemented as follows.  Each particle and trap is assigned an
``internal clock" starting at time $t=0$ according to their waiting time
probability distributions. One particular trap, or the particle, will be
the first to take a step, left or right with equal probability ($1/2$).
We check if trapping of the particle occurs as a result.  If it does, we
stop the dynamics, record the time, and generate a new ensemble of traps
plus one particle. If it does not, we continue the dynamics by observing
the very next trap or particle that takes a step. Again, if trapping
occurs, the time is recorded and the dynamics stopped; if not, the walk
continues.  We also define a maximal time threshold (dictated by our
computational resources) at which we stop the dynamics.

In order to collect enough statistics we have run a large number of
realizations (ensembles of traps and particle) of the dynamics, typically on
the order of $50\,000$.  A $64$-bit congruential random number generator was
used througout the program~\cite{sokal}.  The output of interest of each
realization is the time when the particle is annhilated.  On the basis of this
observable we construct the integrated probability distribution of particle
survival. The statistical errors have been computed using the jackknife
procedure~\cite{jackknife}.

\begin{figure}
\begin{center}
\resizebox{0.8\columnwidth}{!}{\includegraphics{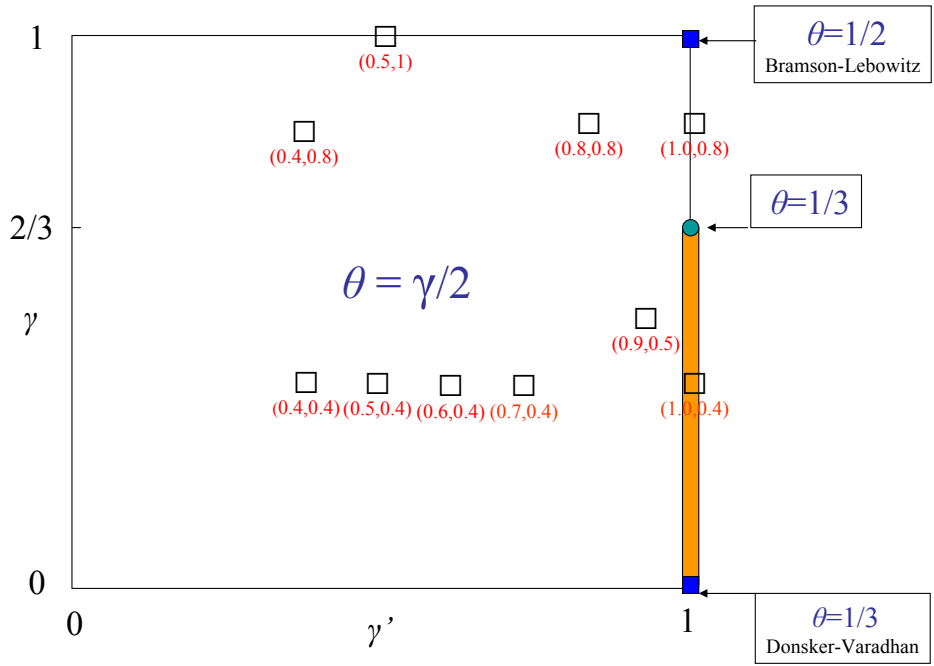}}
\end{center}
\caption{$\gamma$-$\gamma^\prime$ parameter space indicating
regions of analytic asymptotic predictions for the exponent $\theta$
and the coefficient $\lambda$ in $P(t)\sim \exp(-\lambda t^\theta)$.
No predictions are available along the thick strip (orange in colour
version) $\gamma^\prime = 1$, $0<\gamma<2/3$. The exponent is predicted
at the end points of the thick strip, the prefactor only at the lower
end point.}
\label{parameters}
\end{figure}

\section{Results}
\label{results}

We will see that our numerical simulations of the problem $A+B\to B$
when both the traps $B$ and the particle $A$ are mobile approach
asymptotic behavior far more rapidly in some regions of parameter
space than
in others.  Figure~\ref{parameters} shows a $\gamma$-$\gamma^\prime$
space in which are indicated regions and points of exponents
$\theta$ predicted analytically.  The thick (orange in colour rendition)
strip on the right (not including its end points) represents the
parameter regime where the upper and lower survival probability bounds
do not coverge asymptotically and hence no predictions (aside from
conjectures) have been made. The prefactor $\lambda$
has also been predicted everywhere except on the thick (orange) strip
and its upper end point, indicated by a circle (green in colour
rendition).
The empty squares in figure~\ref{parameters} indicate the pairs
($\gamma^\prime, \gamma$) where we have carried out numerical simulations.
All of our results for the apparent exponent $\theta$ obtained from the
simulations for these points are summarized in table~\ref{table}, and a
number of them are subsequently exhibited in figures.

\begin{table}
\caption{Numerical simulation results for the apparent exponent $\theta$
in the survival probability of the particle $A$. The first five sets of
results are for the parameter regime $\gamma < \gamma'$. The next four
sets are for $\gamma \geq \gamma'$. Only the last set is in the regime
$0<\gamma<2/3$, $\gamma'=1$ for which we have no bounding results. In addition
we show the $\chi^2/\mathrm{d.o.f}$ of the fit - $\mathrm{d.o.f}$ being the
number of degrees of freedom of the fit-, the number of realizations performed,
$N_R$, and the logarithm of the minimum value of time used on each fit,
$\ln(t_\mathrm{min}^\mathrm{fit})$. The statistical error in $\theta$ (one
standard deviation) has been computed using a jackknife procedure.}
\begin{indented}
\item[]\begin{tabular}{@{} l l  l  l l l l l}
\br
Traps($\gamma$)&Particle($\gamma^\prime$) & $\rho$
& $\theta$ &$\gamma/2$ & $\ln(t_\mathrm{min}^\mathrm{fit})$ & $\chi^2/\mathrm{d.o.f}$ & $N_R$\\
\mr
&&\centre{3}{$\gamma < \gamma'$}\\
\ns
&&\crule{3}\\
$0.4$ & $0.5$ & $0.01$ & $0.146(7)$ & $0.2$ & $27.2$ & $0.58$ & 460000 \\
      &       & $0.1$ & $0.175(3)$ & $0.2$ & 14.47 & 0.37 & 511733 \\
$0.4$ & $0.6$ & $0.01$ & $0.121(4)$ & $0.2$ & 25.12 & 0.26 & 2050560 \\
      &       & $0.1$ & $0.155(4)$ & $0.2$ &15.74& 1.05& 1398048 \\
$0.4$ & $0.7$ & $0.1$ & $0.093(5)$ & $0.2$ & 24.8 & 0.81 & 6908894 \\
      &       & $0.5$ & $0.190(4)$ & $0.2$ & 8.17 & 1.04 & 2106963 \\
$0.5$ & $0.9$ & $0.01$ & $0.124(2)$ & $0.25$ & 18.5 & 0.6 & 38523723  \\
      &       & $0.1$  & $0.176(4)$ & $0.25$ & 11.9 & 1.08 & 2285460 \\
      &       & $0.5$  & $0.262(6)$ & $0.25$ & 6.25 & 0.93 & 789940  \\
$0.8$ & $1.0$ & $0.01$ & $0.423(2)$ & $0.4$ & 12 & 1.00 &  13626177  \\
      &       & $0.1$   & $0.430(5)$ & $0.4$ &3.55 & 0.83 & 618793\\
&&\crule{3}\\
&&\centre{3}{$\gamma \geq \gamma'$}\\
\ns
&&\crule{3}\\
$0.4$ & $0.4$ & $0.1$ & $0.186(2)$ & $0.2$ & 15.2 & 0.43 & 548122 \\
      &       & $0.5$ & $0.210(2)$ & $0.2$ & 6.47 & 1.05 & 425840\\
$0.8$ & $0.4$ & $0.01$ & $0.396(4)$ & $0.4$ & 11.5 & 0.4 & 111208 \\
      &       & $0.1$  & $0.378(2)$ & $0.4$ & 5.58 & 0.7 & 225796 \\
$0.8$ & $0.8$ & $0.01$ & $0.374(2)$ & $0.4$ & 12.14 & 1.03 & 325275 \\
      &       & $0.1$  & $0.361(3)$ & $0.4$ & 8.64  & 1.03 & 618793 \\
$1.0$ & $0.5$ & $0.01$ & $0.501(2)$ & $0.5$ & 7.5 &  0.5   & 3159970  \\
&&\crule{3}\\
&&\centre{3}{$0<\gamma<2/3,~\gamma'=1$}\\
\ns
&&\crule{3}\\
$0.4$ & $1.0$ & $0.01$ & $0.427(6)$ & $0.2$ & 10 & 1.06 & 11952\\
      &       & $0.1$ & $0.433(7)$  & $0.2$ & 5  & 1.05 & 7740 \\
      &       & $0.5$ & $0.474(7)$  & $0.2$ & 3  & 1.0  &  10000 \\
\br
\end{tabular}
\end{indented}
\label{table}
\end{table}

\begin{figure}
\begin{center}
\resizebox{0.7\columnwidth}{!}{\includegraphics{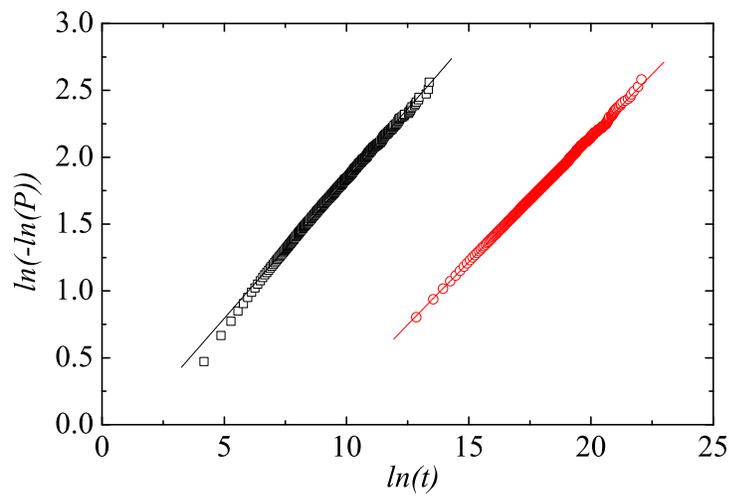}}
\end{center}
\caption{Simulation results for $\gamma=\gamma'=0.4$.
  The left set of data (black in colour version) corresponds to a trap density
  $\rho=0.5$ and a slope of $0.210$; the right set (red in the colour version)
  to $\rho=0.1$ and a slope of $0.186$.  The asymptotic prediction for the
  slope is $\gamma/2=0.2$. The error bars of the points are smaller than the
  size of the symbols (the same in figures ~\ref{fig3} to \ref{fig8}).}
\label{fig2}
\end{figure}

\begin{figure}
\begin{center}
\resizebox{0.7\columnwidth}{!}{\includegraphics{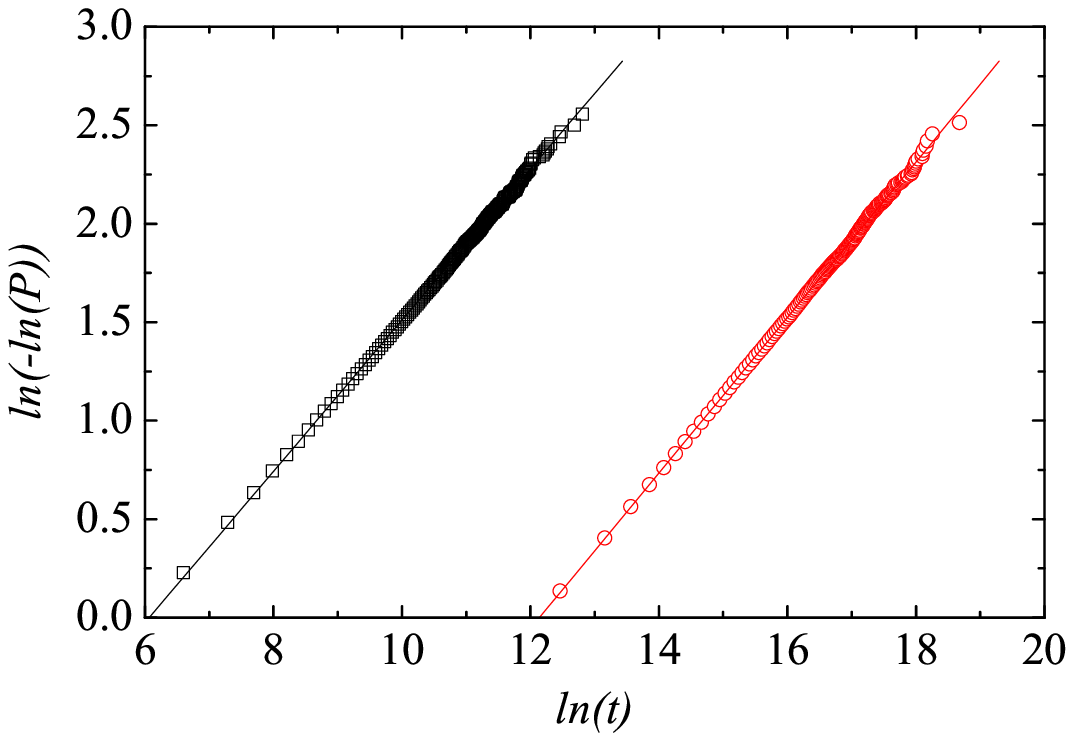}}
\end{center}
\caption{Simulation results for $\gamma=0.8$ and $\gamma'=0.4$.
The left set of data (black) corresponds to a
trap density $\rho=0.1$
and a slope of $0.378$; the right set (red) to $\rho=0.01$ and a slope of
$0.396$.  The asymptotic prediction for the slope is $\gamma/2=0.4$.}
\label{fig3}
\end{figure}

\begin{figure}
\begin{center}
\resizebox{0.7\columnwidth}{!}{\includegraphics{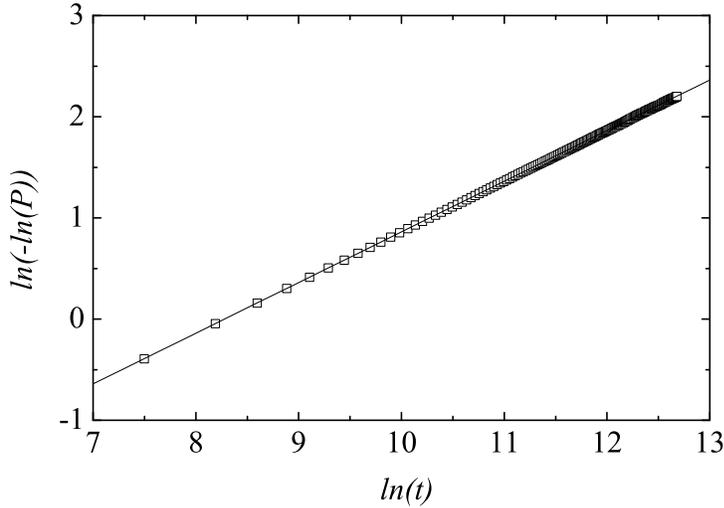}}
\end{center}
\caption{Simulation results for $\gamma=1$ and $\gamma'=0.5$.
Only one set of data points is shown, for $\rho=0.01$.  The slope is
$0.501$. The asymptotic prediction for the slope is $\gamma/2=0.5$.}
\label{fig4}
\end{figure}

The table leads to a number of broad conclusions,
starting with the assertion that in the parameter regime
$\gamma\geq \gamma^\prime$ we have been able to reach the asymptotic
exponent $\theta$, and that this exponent agrees with the theoretical
asymptotic prediction. In this regime
the slope we identify as $\theta$ is indeed insensitive to trap
density changes and close to the asymptotic
value $\gamma/2$.  We have listed four sets of results
for $\gamma\geq \gamma^\prime$,
and exhibit three of them explicitly in figures~\ref{fig2}, \ref{fig3},
and \ref{fig4}.  We see that
the time to achieve asymptotic behavior increases drastically
with decreasing $\gamma$ for a given $\rho$.  For
example, for the same $\rho$ in figures~\ref{fig2} and \ref{fig3}, we
need to go to $\ln(t) > 20$ or so for $\gamma=0.4$ but only to
$\ln(t) > 10$ or so for $\gamma=0.8$.  For a given pair
of parameters $(\gamma,~\gamma')$, the time needed to reach asymptotic
behavior is of course greater when the density of traps is lower.
Note also that this time seems insensitive to the value of $\gamma'$:
for the case in the table not shown in a figure ($\gamma=\gamma'=0.8$),
the abscissa $\ln(t)$ covers the same range for the same trap densities
as those in figure~\ref{fig3}.

\begin{figure}
\begin{center}
\resizebox{0.7\columnwidth}{!}{\includegraphics{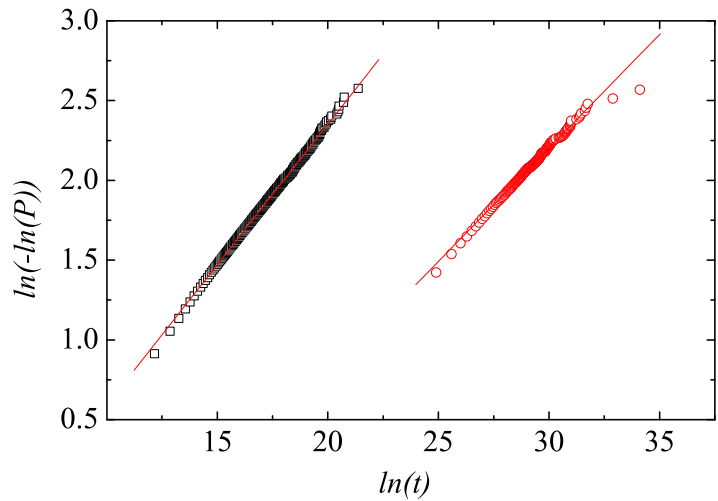}}
\end{center}
\caption{Simulation results for $\gamma=0.4$ and $\gamma'=0.5$.
The left set of data (black in colour version) corresponds to a trap
density $\rho=0.1$ and a slope of $0.175$; the right set (red in
the colour version) to $\rho=0.01$ and a slope of $0.146$.
The asymptotic prediction for the slope is $\gamma/2=0.2$.}
\label{fig5}
\end{figure}

\begin{figure}
\begin{center}
\resizebox{0.7\columnwidth}{!}{\includegraphics{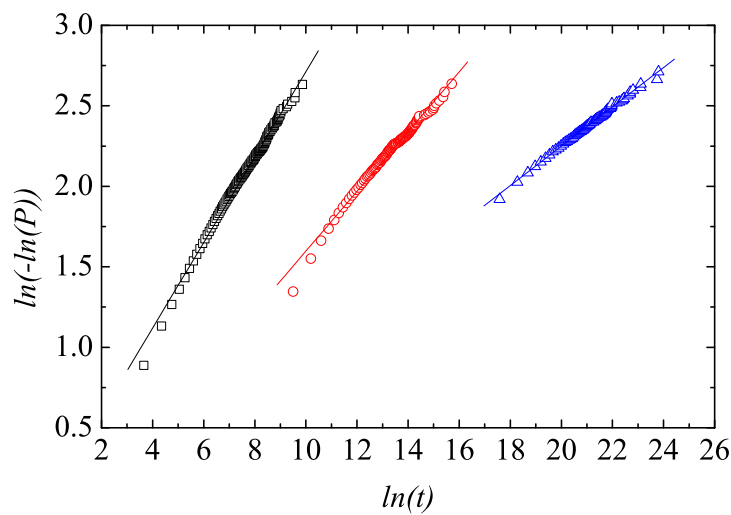}}
\end{center}
\caption{Simulation results for $\gamma=0.5$ and $\gamma'=0.9$.
The left set of data (black) corresponds to a trap
density $\rho=0.5$ and a slope of $0.262$; the middle set (red)
to $\rho=0.1$ and a slope of $0.176$; the right set (blue) to
$\rho=0.01$ and a slope of $0.124$.
The asymptotic prediction for the slope is $\gamma/2=0.25$.}
\label{fig6}
\end{figure}

\begin{figure}
\begin{center}
\resizebox{0.7\columnwidth}{!}{\includegraphics{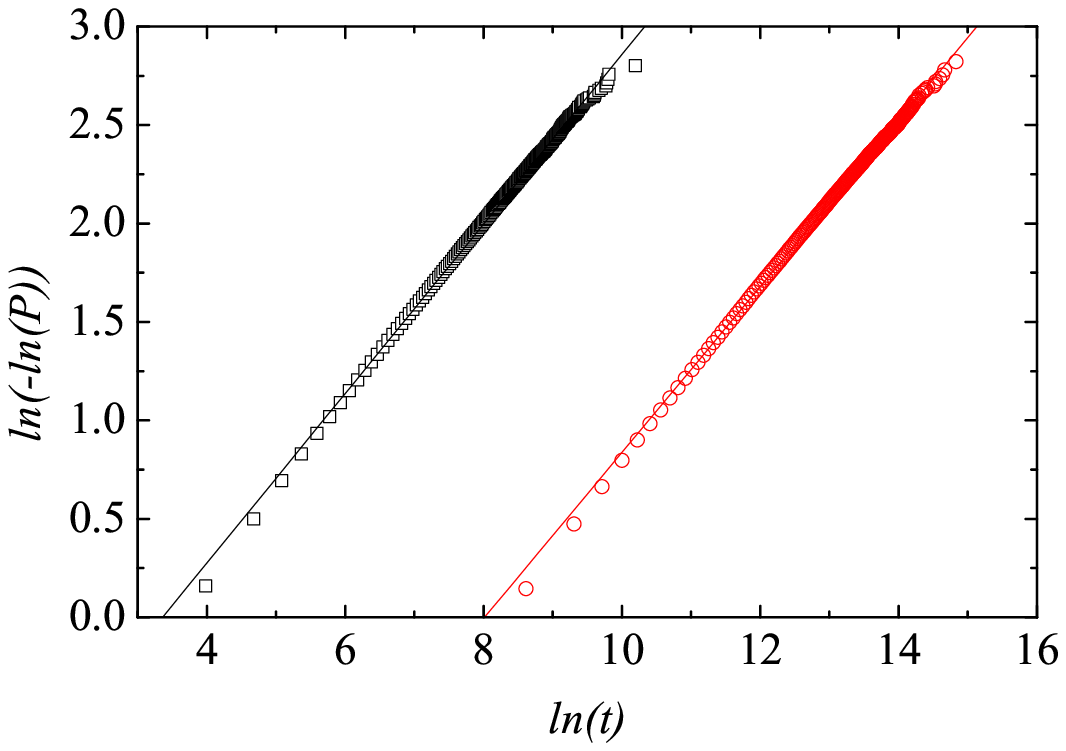}}
\end{center}
\caption{Simulation results for $\gamma=0.8$ and $\gamma'=1$.
The left set of data (black) corresponds to a trap
density $\rho=0.1$ and a slope of $0.430$; the right set (red)
to $\rho=0.01$ and a slope of $0.421$.
The asymptotic prediction for the slope is $\gamma/2=0.4$.}
\label{fig7}
\end{figure}

The situation is more complicated and less satisfactory for
$\gamma  < \gamma'$.  This is seen into the first five sets of results
in table~\ref{table}.  Although (as we will see in the figures) a clear
slope can be read off the simulation data for
$\ln[-\ln P(t)]$, this slope is not independent of $\rho$ in most cases,
nor does it yet satisfactorily approach the theoretically predicted
asymptotic value.  We conclude that in this parameter regime
asymptotia requires a much longer time than we can reasonably simulate,
that is, when the particle moves ``more easily" than the traps,
it takes the system a much longer time to behave as it would if
the particle were simply sitting still.  Nevertheless, within this range of
parameters there are some statements of ``quality" than can be made, as
shown in figures~\ref{fig5}, \ref{fig6}, and \ref{fig7}.   The most
salient point seems to be that asymptotia is reached more readily when
both $\gamma$ and $\gamma'$ are closer to the diffusive case \emph{and}
closer to one another.  Thus the results in figure~\ref{fig7} are
satisfactory (i.e., essentially independent of $\rho$ and in fair
agreement with the asymptotic slope) within a reasonable time range.  If
$\gamma'$ is close to unity but $\gamma$ is too small, it is clearly
difficult to reach asymptotia, as evidenced in the results of
figure~\ref{fig6}.   If both $\gamma$ and $\gamma'$ are small then even
going to extraordinarily long times as shown in figure~\ref{fig5} is not
sufficient.

\begin{figure}
\begin{center}
\resizebox{0.7\columnwidth}{!}{\includegraphics{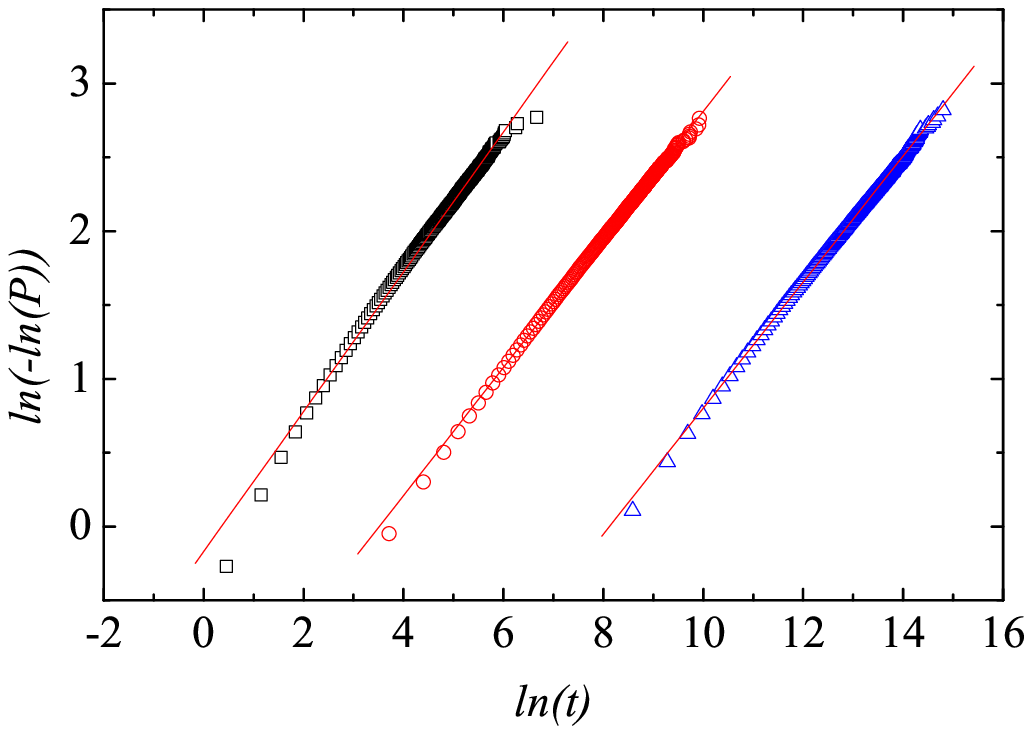}}
\end{center}
\caption{Simulation results for $\gamma=0.4$ and $\gamma'=1$.
The left set of data (black) corresponds to a trap
density $\rho=0.5$ and a slope of $0.474$; the middle set (red)
to $\rho=0.1$ and a slope of $0.433$; the right set (blue) to
$\rho=0.01$ and a slope of $0.427$.  There is no asymptotic prediction
for the slope in this regime, but two conjectures
would be that it might be $1/3$ or $\gamma/2=0.2$.}
\label{fig8}
\end{figure}

Finally, it is apparent that the simulation results for the ``problem
case" $0 < \gamma < 2/3$, $\gamma'=1$ lead to a slope that seems
well defined (see figure~\ref{fig8}) and relatively
insensitive to $\rho$. It is therefore tempting to relate this slope
to asymptotic behavior. However, the value of the slope does not confirm
either of the conjectures put forther earlier, namely, that it is
perhaps equal to $\gamma/2$ or to $1/3$.  The slope is not particularly
close to either of these values. On the other hand, the observation does
not necessarily disprove the conjectures, since it is not clear that
asymptotia has been reached.  In fact, the observed slope is closer to
that expected for the \emph{short-time} behavior of a normally diffusing
particle surrounded by stationary traps, $\theta=1/2$.
Thus the question of the asymptotic behavior in this regime is still open.

We have thus found up to this point that our numerical simulations are
able to confirm the asymptotic prediction for the survival probability
of a particle $A$ characterized by (sub)diffusion exponent $\gamma'$
surrounded by traps $B$ characterized by exponent $\gamma$
provided that
$\gamma < \gamma'$, but that it is difficult to do so
for $\gamma\geq \gamma'$.  For the regime $0<\gamma<2/3,~\gamma'=1$ we
have no asymptotic theory, and the numerical results do not
inform us about the validity of conjectured behaviors.

\begin{figure}
\begin{center}
\resizebox{0.7\columnwidth}{!}{\includegraphics{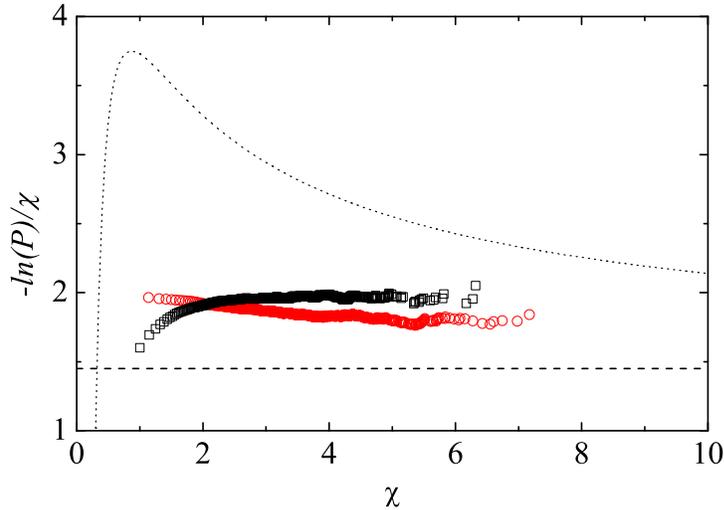}}
\end{center}
\caption{Simulation results for the exponential prefactor $\lambda$
in equation~(\ref{expression}) for $\gamma=\gamma'=0.4$.  Broken line:
upper bound; Dotted curve: lower bound.  Data points
are for $\rho=0.1$ (red, lower set) and $\rho=0.5$ (black, upper set).
}
\label{fig9}
\end{figure}

\begin{figure}
\begin{center}
\resizebox{0.7\columnwidth}{!}{\includegraphics{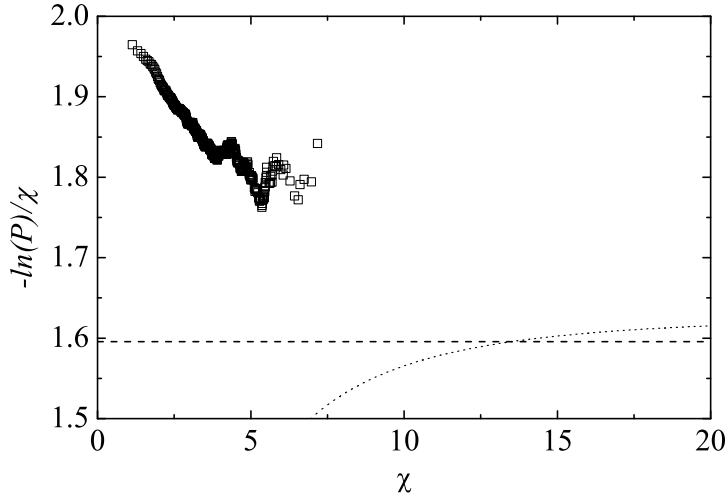}}
\end{center}
\caption{Simulation results for the exponential prefactor $\lambda$
in equation~(\ref{expression}) for $\gamma=1$ and $\gamma'=0.5$.
Broken line: upper bound; Dotted curve: lower bound.  Data points
are for $\rho=0.01$.}
\label{fig10}
\end{figure}

Having determined the exponent $\theta$ in the survival probability
expression~(\ref{expression}) in some parameter regimes, it remains to
explore whether we can numerically determine, or at least bound,
the exponential prefactor $\lambda$.  This turns out to be difficult.  In
figure~\ref{fig9} we show our simulation results for the case
$\gamma=\gamma'=0.4$ of figure~\ref{fig2}.  We also show the upper bound,
which is exactly the asymptotic prediction, cf. compare
equations~(\ref{upperresult}) and (\ref{Ptgral}), and so appears as a
straight (broken) line in the figure.   It lies in the lower part of the
figure because of the minus sign in the ordinate. The lower bound of
equation~(\ref{eq:bound1}), which only approaches the upper bound
asymptotically, is shown as the dotted curve. The simulation results are
for $\rho=0.1$ and $\rho=0.5$, and fall between the bounds.  However,
we would have to go to times far longer than we are able to in order to
ascertain the asymptotic prediction.   A similar figure, but for only one
concentration, $\rho=0.01$, is shown in figure~\ref{fig10} for
$\gamma=1$ and $\gamma'=0.5$.  Here the upper and lower bounds do not
even find their rightful relative placements until a time far beyond our
simulation capabilities, although the simulation results at least point
in the right direction.  It is, in any case, clearly very difficult to
determine the prefactor $\lambda$ and even to ascertain that it is
properly bounded by the theory.

\section{Recap}
\label{conclusions}
In this paper we have made an attempt to assess the validity of the
asymptotic predictions for the survival probability of a (sub)diffusive
particle $A$ characterized by exponent $\gamma'$ surrounded by
(sub)diffusive traps of density $\rho$ characterized by exponent
$\gamma$.  The prediction is arrived at by obtaining an upper and a
lower bound to the survival probability that in most parameter regimes
converge to one another~\cite{our1,our2}.
This asymptotic survival probability in fact
turns out to be exactly the upper bound, which is calculated under the
assumption that particle $A$ remains still. It is thus the case that in
the parameter regime where this prediction is valid it eventually makes
no difference whether or not particle $A$ moves;  the asymptotic survival
probability is entirle determined by the motion of the traps.  However,
when $\gamma'=1$ and the traps are ``too slow" ($0<\gamma< 2/3$), the
bounds no longer converge even asymptotically, and this approach does
not lead to a prediction.  In other words, it is no longer evident that
the motion of the particle does not matter. We have proposed two
conjectures for this regime.  One is that in fact the motion of the
particle does not matter, as before, but our numerical results do not
seem to support this assumption.  The other relies on the fact that
for a diffusive particle we know something about the asymptotic survival
probability at the two extreme points of this interval, namely at
$\gamma=0$ (when the traps are stationary) and at $\gamma=2/3$.  In both
of these cases the survival probability decays as $P(t)\sim \exp\left(
-\lambda t^{1/3}\right)$ (with $\lambda$ known for the former but not
for the latter), and so one might conjecture a $t^{1/3}$ dependence in
the unknown range.  However, this conjecture could not be verified
either.

In the regimes where there is an asymptotic prediction, we are able to
verify it quite clearly when $\gamma \ge \gamma'$, that is, when the
particle moves more slowly or at the same pace than the traps.  Again, the results indicate
that the particle could just as well sit still to reach the same
asymptotic survival probability as it does when it moves.  Also, the
``time to asymptotia" is insensitive to the value of $\gamma'$, but it
is shorter when $\gamma$ is larger and when the density of
traps is higher.   We also tested our ability to predict the asymptotic
exponential prefactor $\lambda$, but find that at best we can show that
it lies between the correct bounds.  At worst, the bounds do not take
their rightful places until times that we can not reach with our
simulations.

When $\gamma < \gamma'$ the situation is far more difficult,
increasingly so with increasing difference between the two exponents. It
would seem to be necessary to go beyond the leading asymptotic term to
thoroughly understand the dynamics for these cases.  This has been done
with some measure of success in the purely diffusive
problem~\cite{anton}.

Our simulation method can not be stretched beyond the times implemented
in this work.  We have been able to answer some questions and ascertain
some predictions, but not others.  To reach the longer times needed to
deal with the questions that we have not been able answer
conclusively will require new simulation optimization methods.  Such
methods have been developed for diffusive particles and
traps~\cite{grassberger}, but their generalization to the subdiffusive
problem does not appear evident.

\ack
This work was partially supported by the Ministerio de Ciencia y
Tecnolog\'{\i}a (Spain) through Grant No. FIS2004-01399, and by the
National Science Foundation under Grant No. PHY-0354937.


\end{document}